\documentclass[a4paper,12pt,oneside]{article}
\usepackage[warn]{mathtext}
\usepackage{amsmath}
\usepackage[T2A]{fontenc}
\usepackage{latexsym}
\usepackage[english]{babel}
\usepackage{graphics}
\usepackage{indentfirst}

\title{Cooperative quasi-Cherenkov radiation}
\author{{\sc S. V. Anishchenko}${}^{1}$\thanks{E--mail: {\tt
      sanishchenko@mail.ru}},{\sc V. G. Baryshevsky}${}^{1}$\thanks{E--mail: {\tt
      bar@inp.bsu.by}}\\
{\small\em ${}^{1}$ Nuclear Problems Institute, Bobruiskaya 11, Minsk 220030, Belarus}}

\date{}
\begin{document}
\maketitle

\begin{abstract}
We study the features of cooperative parametric (quasi-Cherenkov) radiation arising when initially unmodulated electron (positron) bunches pass through a crystal (natural or artificial) under the conditions of dynamical diffraction of electromagnetic waves in the presence of shot noise. A detailed numerical analysis is given for cooperative THz radiation in artificial crystals. The radiation intensity above 200~MW$/$cm$^2$ is obtained in simulations.

In two- and three-wave diffraction cases, the peak intensity of cooperative radiation emitted at small and large angles to particle velocity is investigated as a function of the particle number in an electron bunch. The peak radiation intensity appeared to increase monotonically until saturation is achieved. At saturation, the shot noise causes strong fluctuations in the intensity of cooperative parametric radiation.

It is shown that the duration of radiation pulses can be much longer than the particle flight time through the crystal. This enables a thorough experimental investigation of the time structure of cooperative parametric radiation generated by electron bunches available with modern accelerators. 

The complicated time structure of cooperative parametric (quasi-Cherenkov) radiation can be observed in artificial (electromagnetic, photonic) crystals in all spectral ranges (X-ray, optical, terahertz, and microwave).

\end{abstract}

\section{Introduction}

The generation of short pulses of electromagnetic radiation is a
primary challenge of modern physics. They find applications in
studying molecular dynamics in biological objects and  charge
transfer in nanoelectronic devices, diagnostics of dense plasma
and radar detection of fast moving objects.

The advances in the generation of short pulses of electromagnetic
radiation in infrared, visible, ultraviolet, and X-ray ranges of
wavelengths  are traditionally associated with the development of
quantum electronic devices --- lasers. Radiation in lasers  is
generated via induced  emission of photons by bound electrons.

Electrovacuum devices, operating in a cooperative regime
~\cite{Bonifacio1990,Korovin2006},  have recently become
considered as an alternative to short-pulse lasers, whose active
medium is formed by electrons bound in atoms and molecules. These
are free electron lasers, cyclotron-resonance masers, and
Cherenkov radiators, whose active medium is formed by initially unmodulated electron
bunches propagating in complex electrodynamical structures
(undulators, corrugated waveguides and others). The feature of the
cooperative operation regime lies in the fact that the radiation
power scales as the squared number of particles in the bunch.
This allows calling this regime "superradiance"  by
analogy with the phenomenon predicted by Dicke in quantum
electronics \cite{Bonifacio1990}. 

It should be noted, that the initial phases of charged particles in the electromagnetic wave are homogeneously distributed. As a result, bremsstrahlung produced by oscillating electrons starts from incoherent sponteneous emission. This is true even if the bunch length is much smaller than the radiation wave length. In contrast to bremsstrahlung, Cherenkov (quasi-Cherenkov) radiation starts from coherent spontaneous emission when such a short-length bunch is injected into a slow-wave structure, i. e. the radiation power is proportional to the squared number of particles. The question arises whether this dependence holds when the bunch length increases.

%


%
This paper considers cooperative radiation emitted by electron bunches
when charged particles pass through crystals (natural or
artificial) under the conditions of dynamical diffraction of
electromagnetic waves.
Note that a detailed analysis of the features of incoherent spontaneous
radiation of electrons passing through crystals in both frequency
\cite{BFU} and time \cite{AnishchenkoBaryshevskyGurinovich2012}
domains has been carried out before. This radiation, emitted at
both large and small angles with respect to the direction of
electron motion, is called the parametric (quasi-Cherenkov) radiation. The problems
of amplification of induced parametric X-ray radiation and microwave (optical) quasi-Cherenkov radiation
 have also been thoroughly studied
in the literature \cite{BaryshevskyFeranchuk1984}, and the
threshold current densities providing lasing in crystals have been
calculated \cite{Baryshevsky2012}. Coherent spontaneous radiation produced by modulated electron bunches in crystals has been analysed in \cite{Baryshevsky1984,Ispirian}.

This paper is arranged as follows: In the beginning, a nonlinear
theory of interaction of relativistic charged particles and the
electromagnetic field in crystals is set forth, followed by the
results of numerical calculations of the parametric radiation
pulse. The dependence of the radiation
intensity on the particle number in an electron bunch and the geometrical
parameters of the system is considered. The appendix outlines the
algorithm used in the simulation.  The feature of the algorithm is
that it is based on the particle-in-cell method \cite{MoreyBirdsall1989}, which enables
studying kinetic phenomena. Let us note that in most of the
existing codes (see, e.g. \cite{Antonsen,Ginzburg}) used
for simulating the interaction of charged particles and a
synchronous wave, the motion of charged particles is considered
within the framework of the hydrodynamic approximation.

\section{Nonlinear theory of cooperative radiation}

A theoretical analysis of radiation can be performed
only by means of a  self-consistent solution of a nonlinear set of
the Newton--Maxwell equations:
\begin{equation}
\label{Newton}
\frac{d\vec p_j}{dt}=q_e\big(\vec E(\vec r_j,t)+\vec v_j\times\vec H(\vec r_j,t)\big),
\end{equation}
\begin{eqnarray}
\label{Maxwell}
 \nabla\times\vec E=-\frac{1}{c}\frac{\partial\vec H}{\partial t},\nonumber\\
 \nabla\times\vec H=\frac{1}{c}\frac{\partial\vec D}{\partial t}+\frac{4\pi}{c}\vec j,\nonumber\\ 
 \nabla\cdot\vec D=4\pi\rho,\nonumber\\
 \nabla\cdot\vec H=0,
\end{eqnarray}
describing the electron motion in the electric $\vec E$ and
magnetic $\vec H$ fields.
Here $\vec j=q_e\sum_j\vec v_j\delta(\vec r-\vec r_j)$ and $\rho=q_e\sum_j\delta(\vec r-\vec r_j)$ are the current
and charge densities, respectively. Since the crystal is a
periodic linear medium with frequency dispersion,  the Fourier
transform of the electric displacement field $\vec D(\vec r,\omega)$
relates to the electric field $\vec E(\vec r,\omega)$ as $D(\vec
r,
\omega)=\big(1+\chi_0(\omega)+\sum\limits_{\vec\tau}2\chi_{\vec\tau}(\omega)\cos(\vec\tau\vec
r)\big)\vec E(\vec r, \omega)$, where the summation is made over
all reciprocal lattice vectors.
The dielectric susceptibilities in natural crystals in the X-ray range and in
grid photonic crystals built from metallic threads are inversely proportional to the frequency
\cite{BaryshevskyGurinovich2006}:
$\chi_{0,\vec\tau}(\omega)=\Omega_{0,\vec\tau}^2/\omega^2$.
(We should underline that, in the case of photonic crystals built from metallic threads, the equality $\chi_{0,\vec\tau}(\omega)=\Omega_{0,\vec\tau}^2/\omega^2$ is valid when a thread radius is much smaller than the radiation wavelength). 
This permits to reduce Maxwell's equations (\ref{Maxwell}) to the equation
of the form:
\begin{equation}
\label{Maxwell2}
\frac{1}{c^2}\frac{\partial^2\vec E}{\partial t^2}+\nabla(\nabla\cdot\vec E)-\Delta\vec E+\frac{\Omega_0^2}{c^2}\vec E+\sum\limits_{\vec\tau}2\frac{\Omega_{\vec\tau}^2}{c^2}\cos(\vec\tau\vec r)\vec E=-\frac{4\pi}{c^2}\frac{\partial\vec j}{\partial t}.
\end{equation}

Let's simplify the equation (\ref{Maxwell2}) for the case when two strong waves
are excited in the crystal: the forward wave and the diffracted
wave (the so-called two-wave diffraction case). The forward wave
(its wave   vector is denoted by $\vec k$) is emitted at small
angles with respect to the particle velocity, while the diffracted
one, having the wave vector $\vec k_\tau=\vec k+\vec\tau$,  is
emitted at large angles to it (Fig.~1). Under the conditions of
dynamical Bragg diffraction, the following relation is fulfilled: $\vec
k_\tau^2\approx \vec k^2\approx\omega^2/c^2$.

\begin{figure}[ht]
\begin{center}
       \resizebox{65mm}{!}{\includegraphics{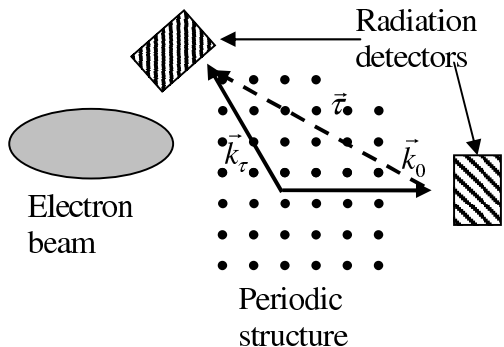}}\\
\caption{The two-wave diffraction geometry.}
\label{Fig.1}
\end{center}
\end{figure}

Let us perform the following  simplifications: First,  we shall
neglect the longitudinal ($\nabla\cdot\vec D\to0$) fields of the
bunch, which is appropriate when the value of the Langmuir
oscillation frequency $\Omega_b$ of the bunch is less than the
values of $\Omega_{0,\tau}$. In this case, the Coulomb forces will
not have an appreciable effect on  electrodynamical properties of the
system. Second, we shall seek for the electric field $\vec E$
using the method of slowly varying amplitudes. Third, we shall
assume that a transversally infinite bunch executes
one-dimensional motion along the $OZ$-axis (this is achieved by
inducing a strong axial magnetic field in the system).

Under the conditions of two-wave diffraction, the field $\vec E$
can be presented as a sum
\begin{equation}
\label{ElectricField}
 \vec E=\vec e_0E_0(z,t)e^{i(\vec k\vec r-\omega t)}+\vec e_\tau E_\tau(z,t) e^{i(\vec k_\tau\vec r-\omega t)},
\end{equation}
where the amplitudes of the forward $E_0$ and diffracted $E_\tau$
waves are slowly varying variables. This means that for the distances
comparable with the wavelength and the  times comparable with the
oscillation period, the values of  $E_0$ and~$E_\tau$ remain
practically the same. Substituting (\ref{ElectricField}) into
(\ref{Newton}) and  (\ref{Maxwell2}) and then averaging them over
the length $l=2j\lambda=2j\pi/k_z$, where $j$ is a natural number,
we obtain
\begin{equation}
\label{Vlasov3}
 \frac{dp_{zj}}{dt}=2q_ee_{0z}Re\Big(E_0e^{i(k_zz_j-\omega t)}\Big),
\end{equation}
\begin{eqnarray}
\label{Maxwell3}
\frac{1}{c}\frac{\partial E_0}{\partial t}+\gamma_0\frac{\partial E_0}{\partial z}+\frac{i\Omega_0^2}{2\omega c}E_0+\frac{i\Omega_\tau^2}{2\omega c}E_\tau=-\frac{2\pi}{c}\int_{z-l/2}^{z+l/2} e_{0z}j_ze^{i(\omega t-k_zz)}dz/l,\nonumber\\ 
\gamma_0=k_z/k,\nonumber\\
\frac{1}{c}\frac{\partial E_\tau}{\partial t}+\gamma_\tau\frac{\partial E_\tau}{\partial z}+\frac{i\Omega_0^2}{2\omega c}E_\tau+\frac{i\Omega_\tau^2}{2\omega c}E_0=-\frac{2\pi}{c}\int_{z-l/2}^{z+l/2} e_{\tau z}j_ze^{i(\omega t-k_{\tau z}z)}dz/l,\nonumber\\
\gamma_\tau=k_{\tau z}/k.
\end{eqnarray}

Now let us complete the set of equations (\ref{Vlasov3}) and
(\ref{Maxwell3}) with boundary conditions (the initial conditions
are reduced to the condition that all values of the fields equal
zero at $t=0$): in the plane $z=0$, let us specify the time
dependence of function $f$ and set the field $E_0$  to zero. In
the case of Bragg diffraction, the boundary condition imposed on
the diffracted wave is reduced to the condition that the field
$E_\tau$ equals zero at $z=L$, while in the case of Laue
diffraction, it equals zero at $z=0$.

The difference between the two diffraction schemes is not merely
kinematic, but radical. In the Bragg case, there
is a synchronous wave moving against the electrons of the beam,
which gives rise to the  internal feedback and absolute
instability.
In Laue diffraction geometry, a backward wave is absent, and as a
result absolute instability does not evolve. 
It may
seem that electromagnetic radiation is not generated. However,
fluctuations of the electron current (the shot noise), which always occur in real beams, are amplified when
the beam enters the  crystal (due to convective instability,
excited in the beam).

In analyzing multiparametric problems, to which the problem of
cooperative parametric (quasi-Cherenkov) radiation refers, it is convenient to write
equations (\ref{Vlasov3}) and (\ref{Maxwell3}) in a dimensionless
form. This procedure enabled transferring the calculation results
from one set of parameters to another. The substitution of
variables $\omega t\to t$, $\omega L/c\to L$,
$mc\omega E_{0,\tau}/q_e\to E_{0,\tau}$
then gives

\begin{equation}
\label{Sim1}
 \frac{dp_{zj}}{dt}=2\theta Re\Big(E_0e^{-i(t-k_zz_j+\phi_j)}\Big),
\end{equation}
\begin{eqnarray}
\label{Sim2}
\frac{\partial E_0}{\partial t}+\gamma_0\frac{\partial E_0}{\partial z}+\frac{i\chi_0}{2}E_0+\frac{i\chi_\tau}{2}E_\tau=-\sum_j\frac{\theta\chi_{bj}}{2} \frac{e^{i(\omega t-k_zz_j+\phi_j)}}{N_l},\nonumber\\
\frac{\partial E_\tau}{\partial t}+\gamma_\tau\frac{\partial E_\tau}{\partial z}+\frac{i\chi_0}{2}E_\tau+\frac{i\chi_\tau}{2}E_0=0.
\end{eqnarray}
Here the quantity $\chi_{bj}=-4\pi q_e^2n_j/m\omega^2$ is
determined at the moment when the $j$th particle enters the
system, $n_j$ is the corresponding  electron density, $\theta$ is the
angle between the particle velocity and the wave vector $\vec k$, $\phi_j$ is the initial phase of the $j$th particle, and $N_l$ is
the number of particles over the length $l$.
The set of equations with boundary conditions contains four
independent parameters: $\chi_{0,\tau}$, $\omega L/c$, $\theta$ that define
the geometry of the system. In addition to these parameters, we
need to specify the beam profile. Let
\begin{equation}
 \chi_b=\chi_{b0}\exp(-z^2/L_b^2),
\end{equation}
where the bunch length $L_b$ is further assumed to be equal
to $0.1L/c$.

Obviously, in the three-diffraction case, the set of equations analogues to (\ref{Sim2}) should be rewritten as follows:
\begin{eqnarray}
\label{Sim3}
\frac{\partial E_1}{\partial t}+\gamma_1\frac{\partial E_1}{\partial z}+\frac{i\chi_0}{2}E_1+\frac{i\chi_\tau}{2}(E_2+E_3)=-\sum_j\frac{\theta\chi_{bj}}{2} \frac{e^{i(t-k_{z1}z_j+\phi_j)}}{N_l},\nonumber\\
\frac{\partial E_2}{\partial t}+\gamma_2\frac{\partial E_2}{\partial z}+\frac{i\chi_0}{2}E_2+\frac{i\chi_\tau}{2}(E_3+E_1)=0,\nonumber\\
\frac{\partial E_3}{\partial t}+\gamma_3\frac{\partial E_3}{\partial z}+\frac{i\chi_0}{2}E_3+\frac{i\chi_\tau}{2}(E_1+E_2)=0.
\end{eqnarray}

\section{Simulation results}
 The characteristic feature of  cooperative pulses is the
 peculiar dependence of the peak power on the number of particles $N_b$ in the
 bunch. When the particles are small in number,  the radiation power monotonically increases until saturation is achieved.

  \begin{figure}[ht]
  \begin{center}
       \resizebox{65mm}{!}{\includegraphics{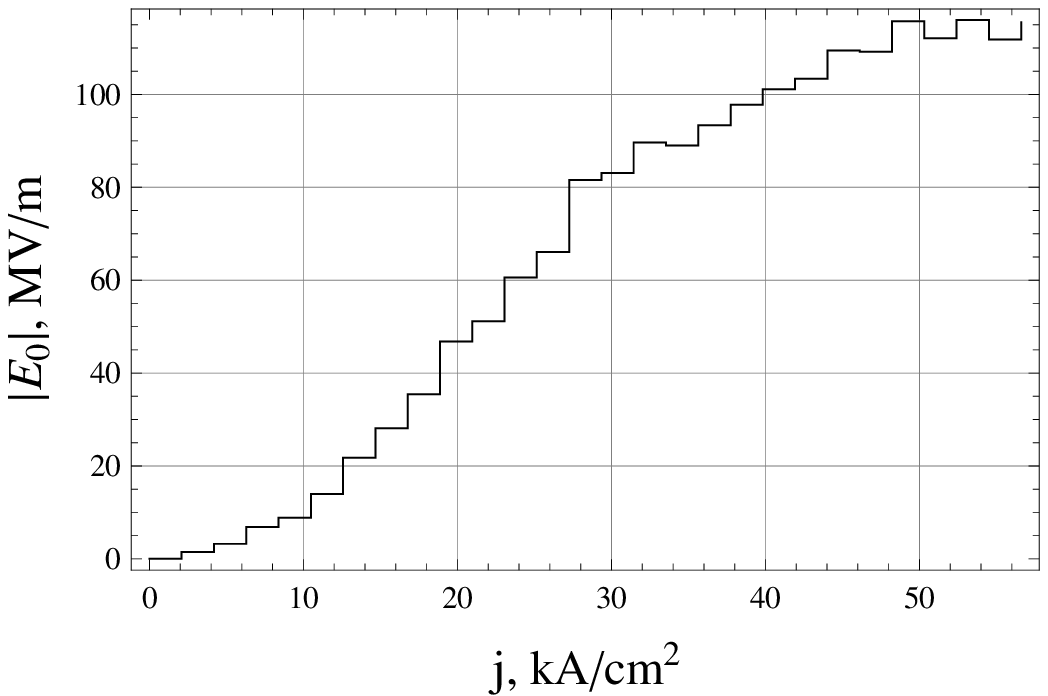}} \resizebox{65mm}{!}{\includegraphics{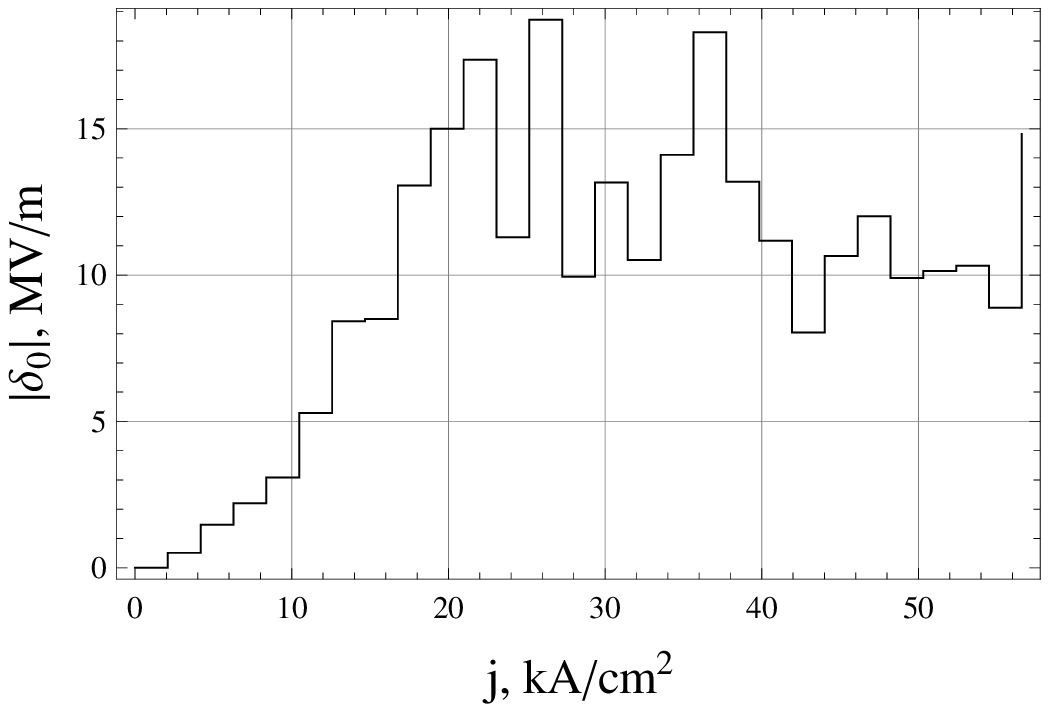}}\\
  \end{center}
    \caption{The Bragg case. Quasi-Cherenkov radiation at small angles to particles' velocities: radiation amplitude (left), amplitude dispersion (right) [$\theta_B=67.5^{o}$, $\nu=0.1$~THz].}
\label{Fig.2}
  \end{figure}
  \begin{figure}[ht]
  \begin{center}
       \resizebox{65mm}{!}{\includegraphics{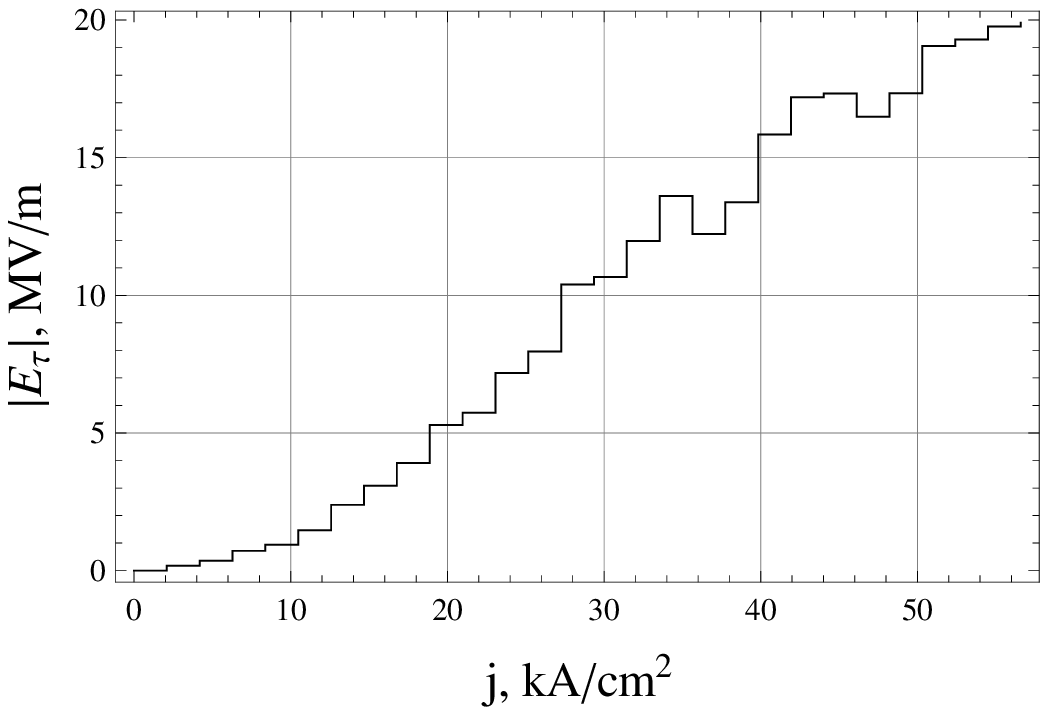}} \resizebox{65mm}{!}{\includegraphics{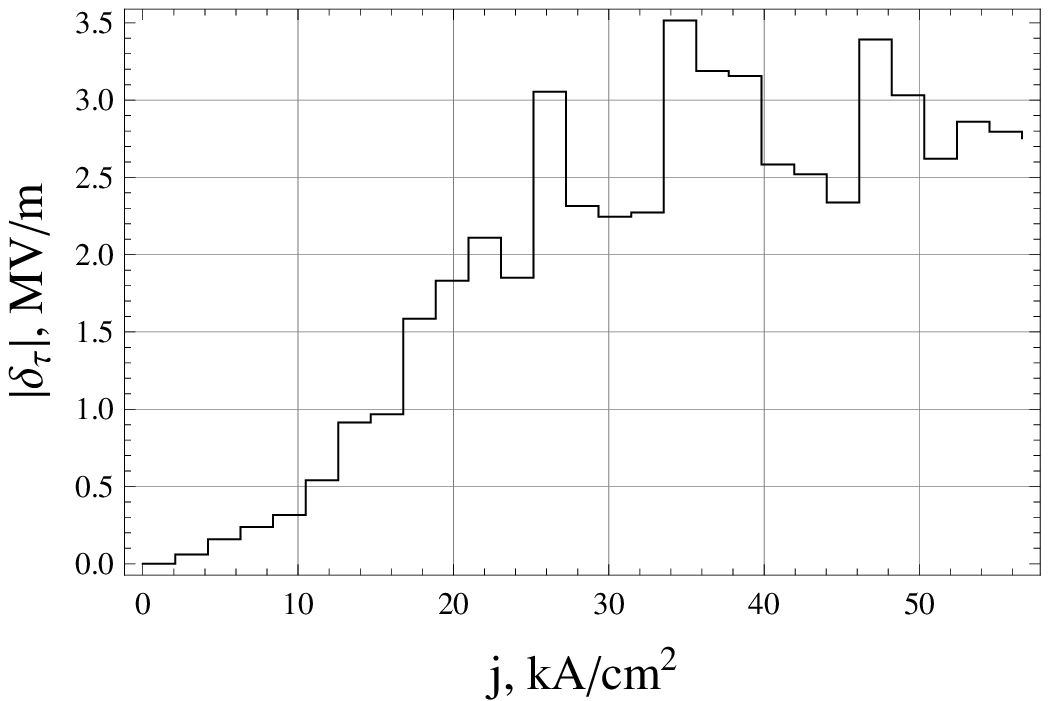}}\\
  \end{center}
    \caption{The Bragg case. Quasi-Cherenkov radiation at large angles to particles' velocities: radiation amplitude (left), amplitude dispersion (right) [$\theta_B=67.5^{o}$, $\nu=0.1$~THz].}
\label{Fig.3}
  \end{figure}
\begin{figure}[ht]
  \begin{center}
       \resizebox{65mm}{!}{\includegraphics{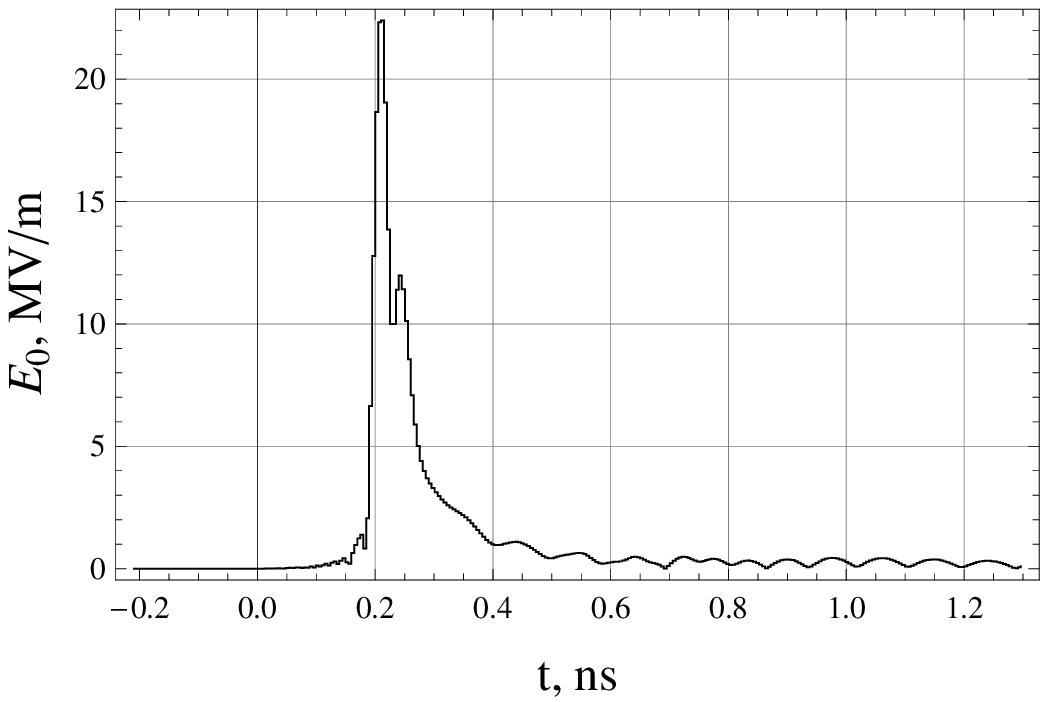}} \resizebox{65mm}{!}{\includegraphics{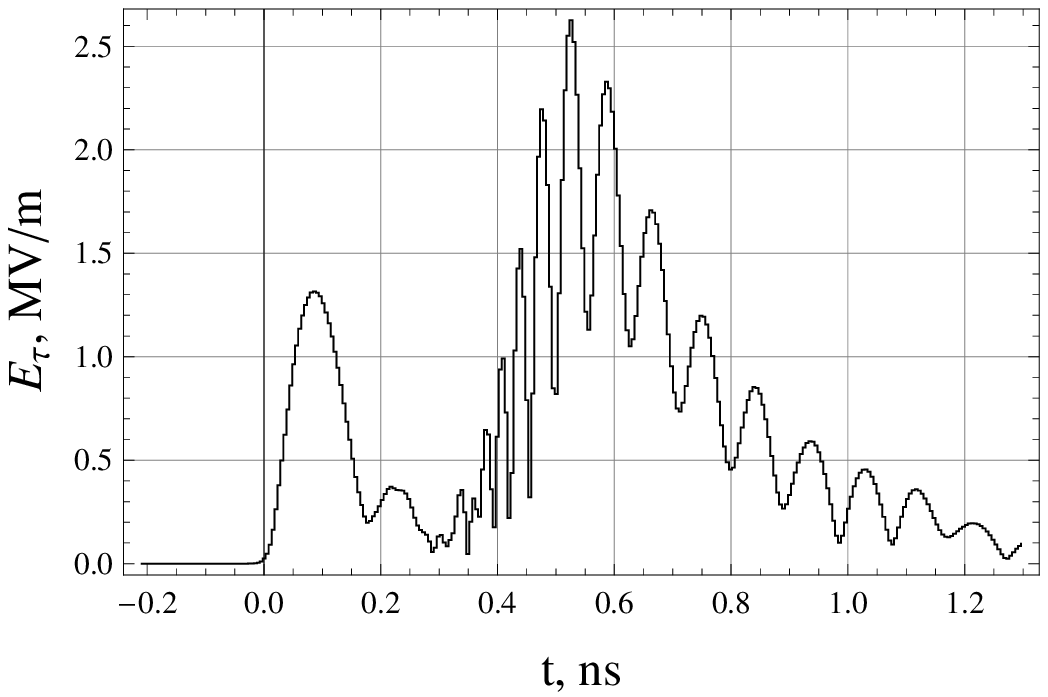}}\\
  \end{center}
    \caption{The Bragg case. Quasi-Cherenkov radiation at small  (left) and (large) angles to particles' velocities [$\theta_B=67.5^{o}$, $j=10$~kA$/$cm$^2$, $\nu=0.1$~THz].}
\label{Fig.4}
  \end{figure}
\begin{figure}[ht]
  \begin{center}
       \resizebox{65mm}{!}{\includegraphics{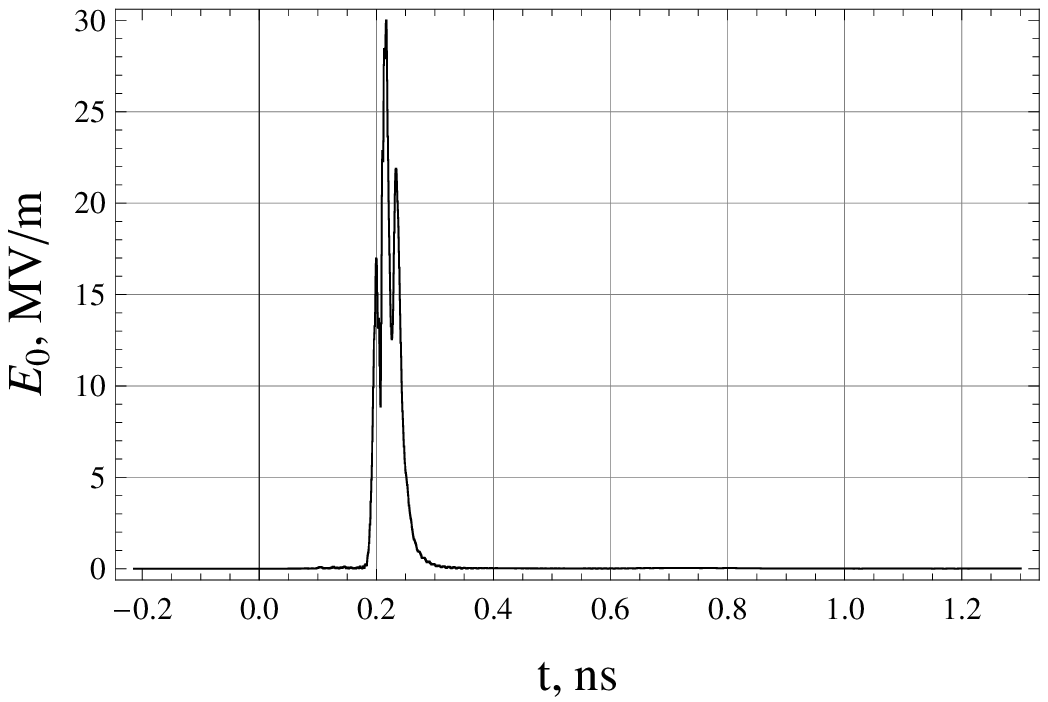}} \resizebox{65mm}{!}{\includegraphics{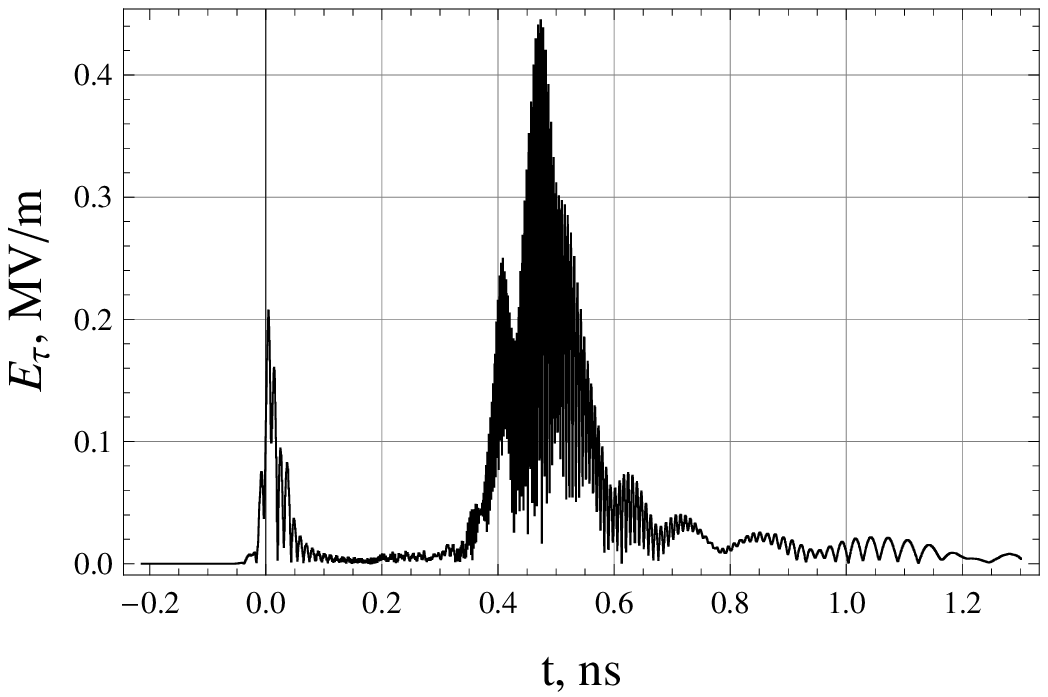}}\\
  \end{center}
    \caption{The Bragg case. Quasi-Cherenkov radiation at small angles to particles' velocities: radiation amplitude (left), amplitude dispersion (right) [$\theta_B=67.5^{o}$, $\nu=1.0$~THz].}
\label{Fig.5}
  \end{figure}

  \begin{figure}[ht]
  \begin{center}
       \resizebox{65mm}{!}{\includegraphics{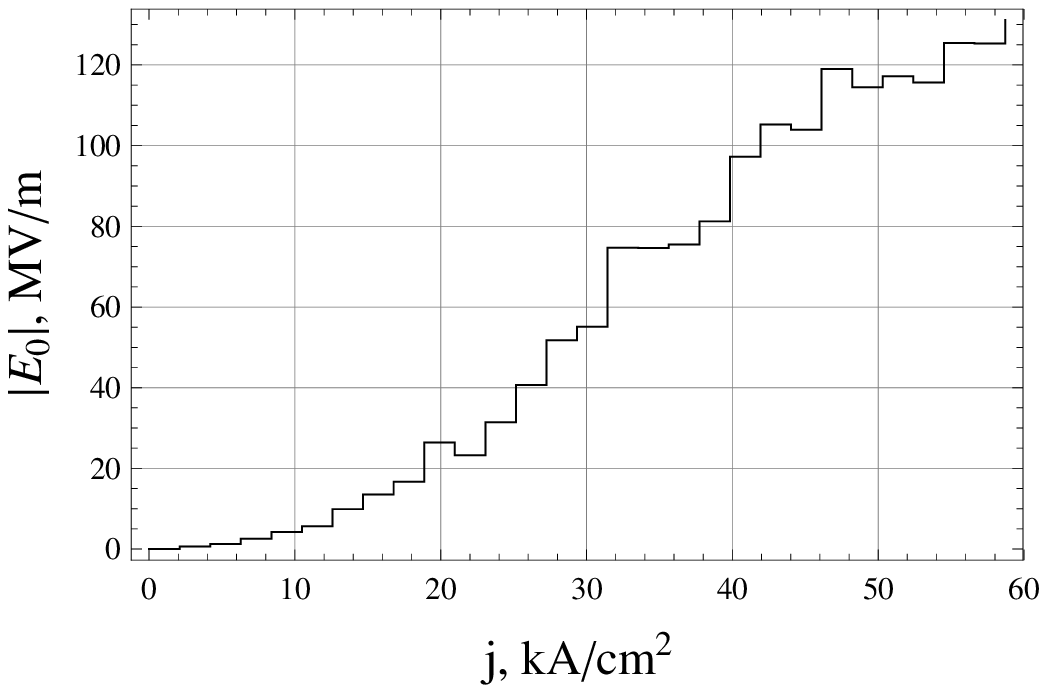}} \resizebox{65mm}{!}{\includegraphics{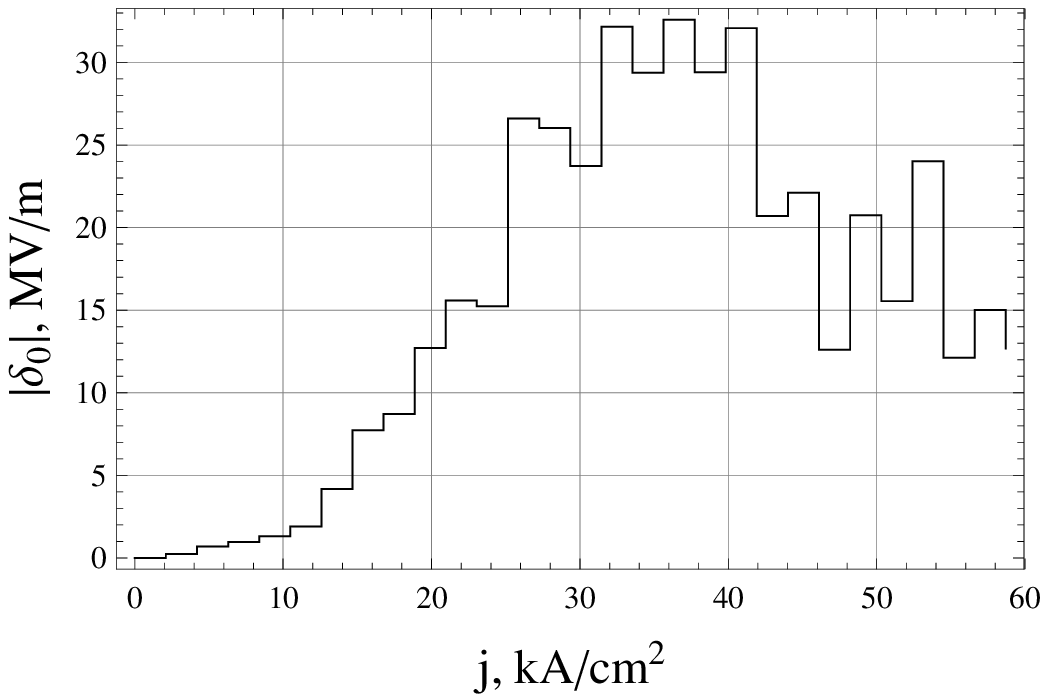}}\\
  \end{center}
    \caption{The Laue case. Quasi-Cherenkov radiation at small angles to particles' velocities: radiation amplitude (left), amplitude dispersion (right) [$\theta_B=22.5^{o}$, $\nu=0.1$~THz].}
\label{Fig.6}
  \end{figure}
  \begin{figure}[ht]
  \begin{center}
       \resizebox{65mm}{!}{\includegraphics{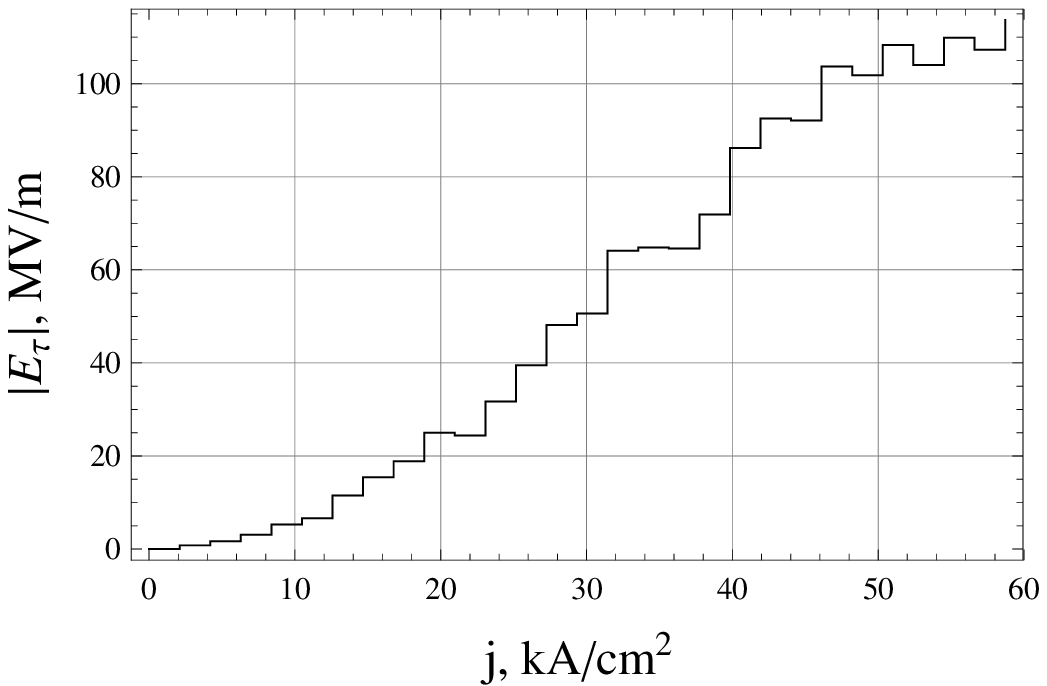}} \resizebox{65mm}{!}{\includegraphics{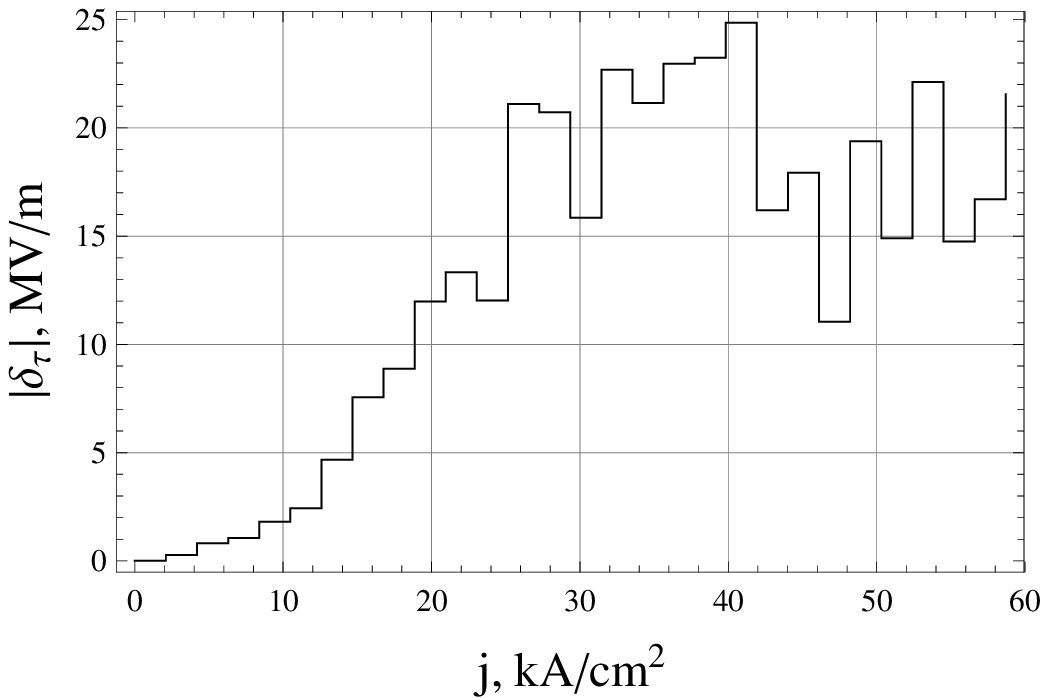}}\\
  \end{center}
    \caption{The Laue case. Quasi-Cherenkov radiation at large angles to particles' velocities: radiation amplitude (left), amplitude dispersion (right) [$\theta_B=22.5^{o}$, $\nu=0.1$~THz].}
\label{Fig.7}
  \end{figure}
\begin{figure}[ht]
  \begin{center}
       \resizebox{65mm}{!}{\includegraphics{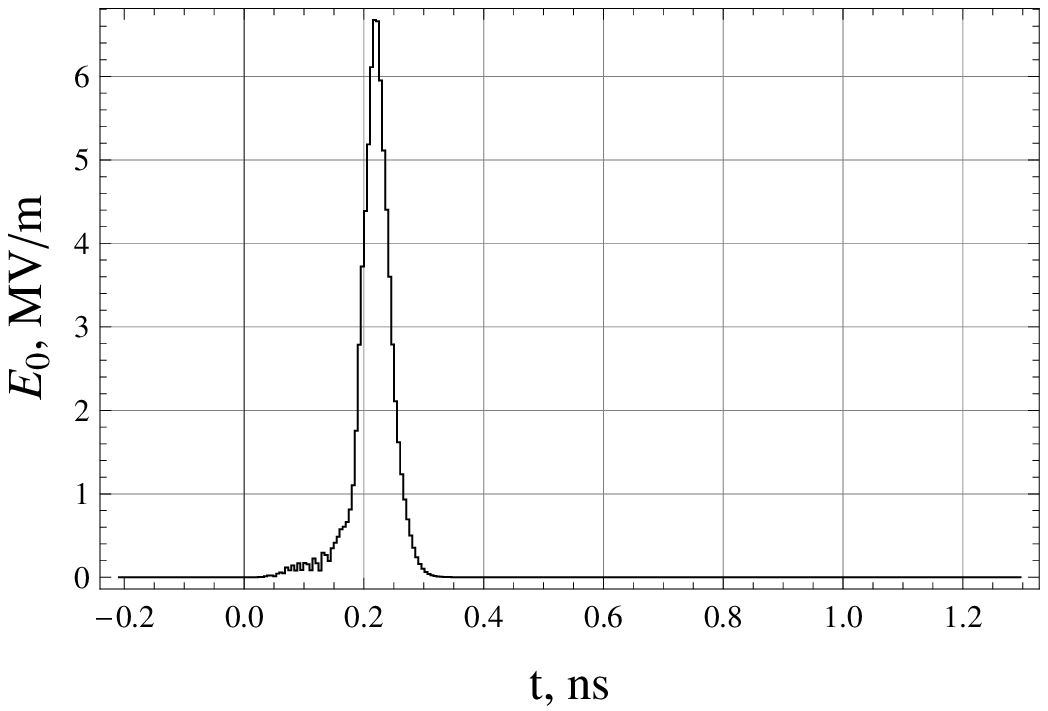}} \resizebox{65mm}{!}{\includegraphics{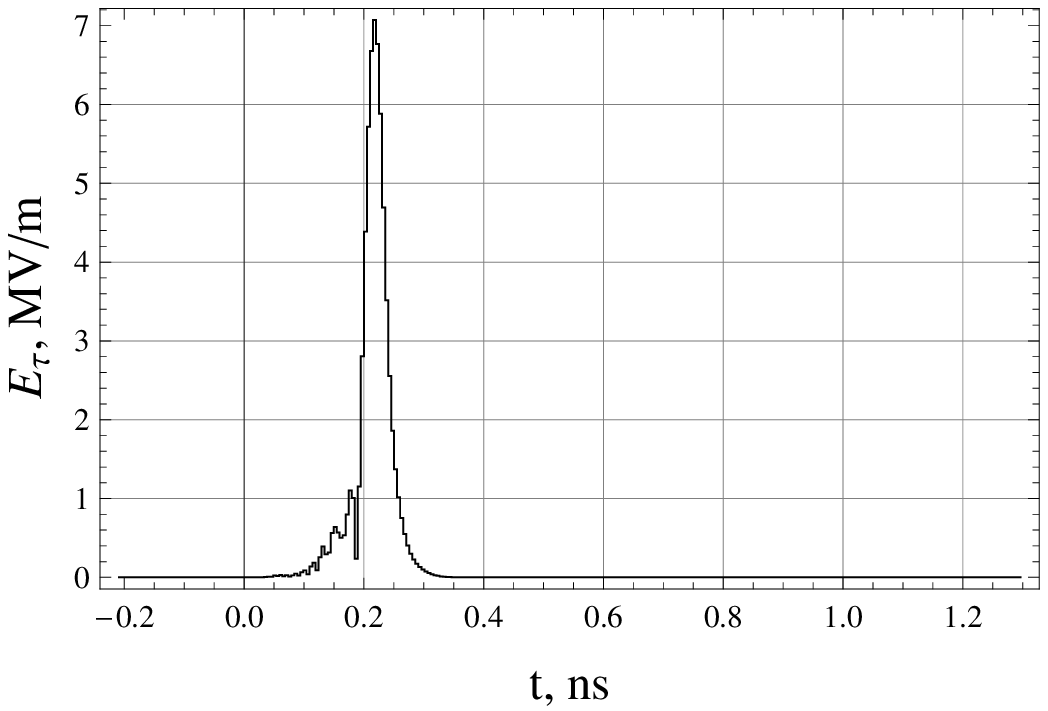}}\\
  \end{center}
    \caption{The Laue case. Quasi-Cherenkov radiation at small  (left) and (large) angles to particles' velocities [$\theta_B=22.5^{o}$, $j=10$~kA$/$cm$^2$, $\nu=0.1$~THz].}
\label{Fig.8}
  \end{figure}

\begin{figure}[ht]
  \begin{center}
       \resizebox{65mm}{!}{\includegraphics{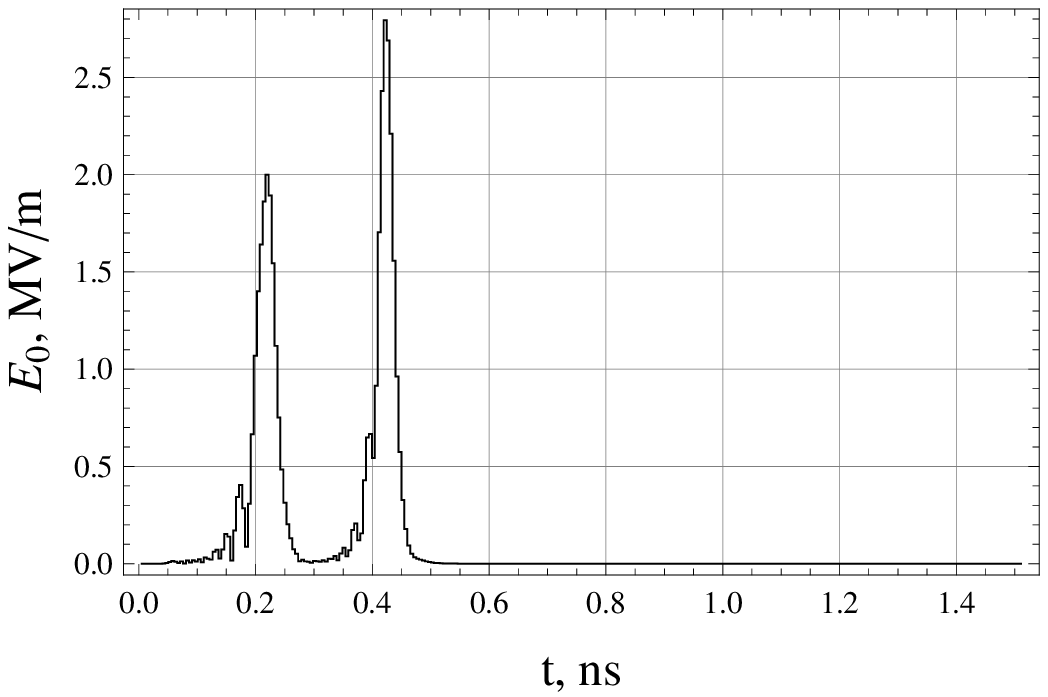}} \resizebox{65mm}{!}{\includegraphics{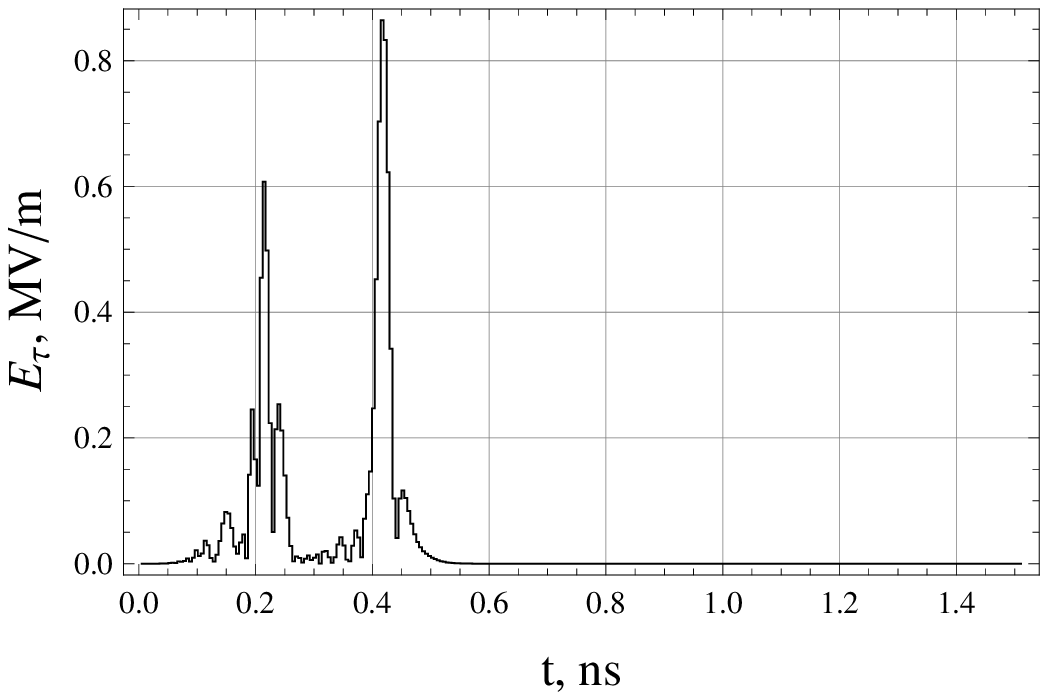}}\\
  \end{center}
    \caption{The Laue case. Quasi-Cherenkov radiation in the absence of shot noise [$\theta_B=22.5^{o}$, $L_b/L=1.0$, $j=10$~kA$/$cm$^2$, $\nu=0.1$~THz].}
\label{Fig.9}
  \end{figure}
  \begin{figure}[ht]
  \begin{center}
       \resizebox{65mm}{!}{\includegraphics{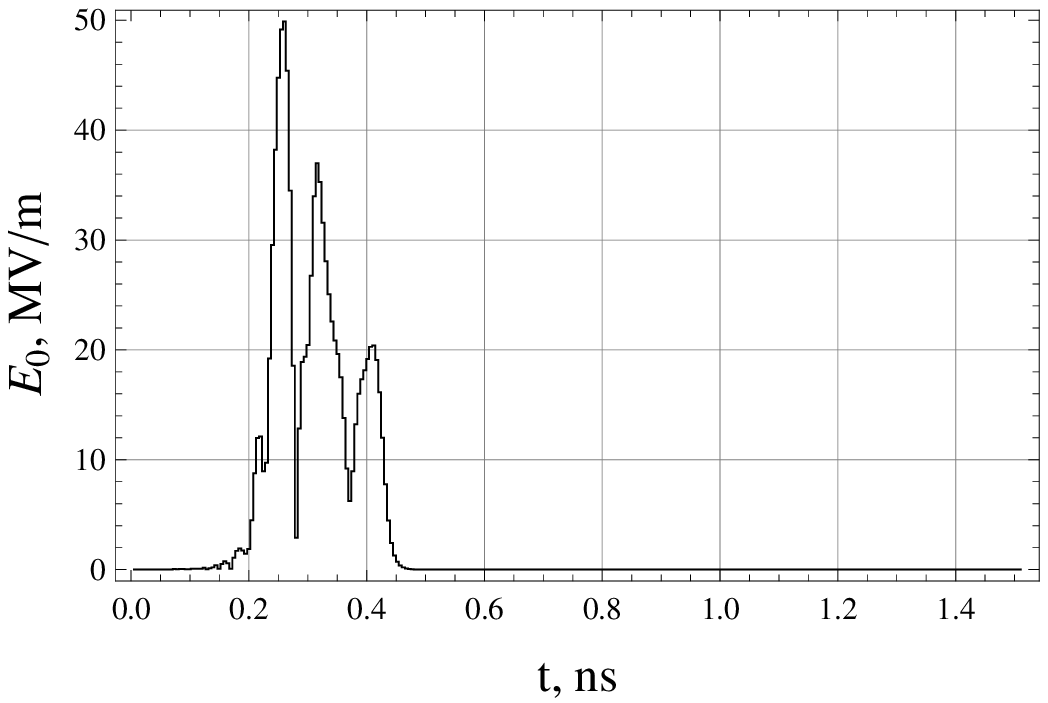}} \resizebox{65mm}{!}{\includegraphics{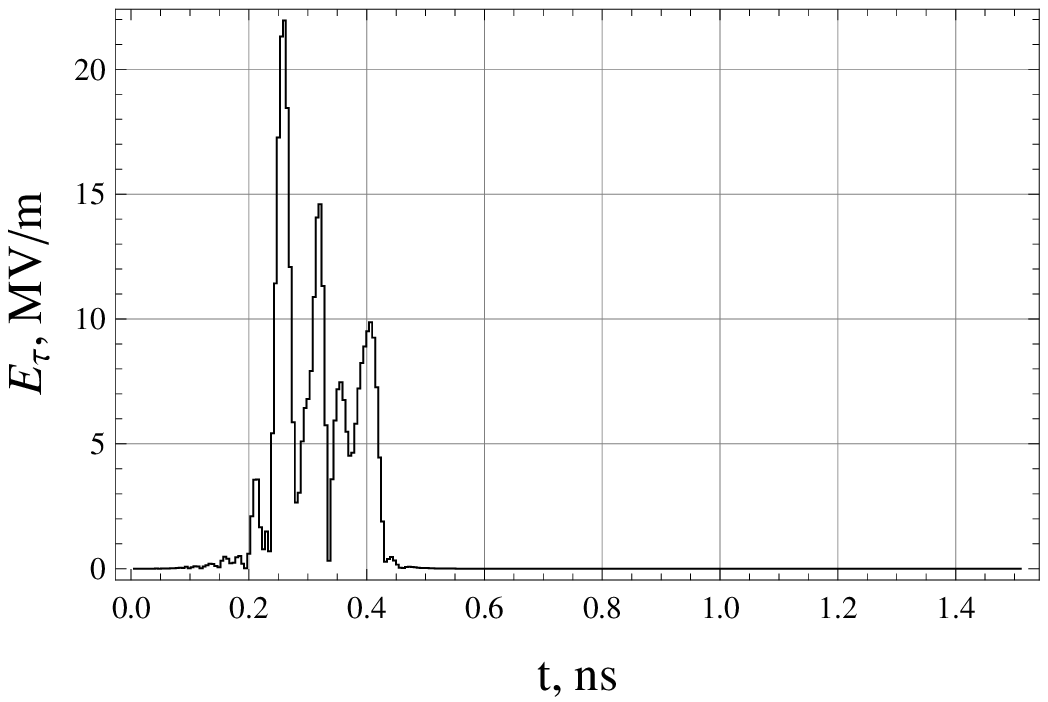}}\\
  \end{center}
    \caption{The Laue case. Quasi-Cherenkov radiation in the presence of shot noise [$\theta_B=22.5^{o}$, $L_b/L=1.0$, $j=10$~kA$/$cm$^2$, $\nu=0.1$~THz]}
\label{Fig.10}
  \end{figure}

Let us see now how the dynamical diffraction of electromagnetic
waves affects  the cooperative radiation in crystals.
How the
Bragg diffraction case is different from the case of Laue
diffraction?

We shall assume that $\gamma=3.0$, $\theta=0.33$~rad, $\chi_0/2=\chi_\tau=0.1$, and $L=6$~cm.

Let us start our consideration  with the Bragg case. In this case, along with the electromagnetic wave emitted in the forward direction, one can observe the electromagnetic wave that is emitted by charged particles in the diffraction direction and leaves the crystal through the bunch entrance surface.

The peak radiation field emitted at small and large angles to particle velocity is investigated as a function of the peak current density $j$. The peak radiation intensity appeared to increase monotonically until saturation is achieved (Fig. 2---3). At saturation, the shot noise causes strong fluctuations in the intensity of cooperative parametric radiation. The amplitute dispersion $\delta_{0,\tau}=\sqrt{<E_{0,\tau}>^2-<E_{0,\tau}^2>}$ is presented on fig. 2 (the brackets $<..>$ denote average values). 

The results of computation (Fig. 4) show that the cooperative radiation emitted at large angles lasts much longer ($t_{rad}\sim0.6$~ns) than the particle flight time through the crystal ($t_p=0.2$~ns), though that is much lower than the radiation intensity emitted in forward direction. 
 We would like to note that the long duration of parametric radiation can be observed in sponteneous processes too \cite{AnishchenkoBaryshevskyGurinovich2012}.

Figures 2---4 correspond to the frequency $\nu=0.1$~THz and current density $j=10$~kA$/$cm$^2$. If we increase $\nu$ in ten times leaving the current density unchanged the peak radiation field will be $E\approx30$~MV$/$m which corresponds to 240~MW$/$cm$^2$ (fig. 5).

Now, let us consider the Laue case. In this case, the electromagnetic waves emitted by charged particles in the forward and diffraction directions leave the crystal through the same surface.
Under Laue diffraction conditions (Fig. 6---8), the pulses of parametric
radiation emitted in forward and diffracted directions have comparable amplitudes and durations. 

We should point out that the shot noise results in strong fluctuations in radiation intensity at saturation. Under Laue diffraction conditions, the shot noise leads to an appreciable change in the pulse form due to the convective
character of instability: in the absence of noise, the generation
occurs only at the ends of the bunch of charged
particles. As a result, the cooperative pulse posses a
two-peak structure (Fig. 9).  The presence of noise leads to an
appreciable change in the pulse form: the interval between the two
pulses is filled with a chaotic signal (Fig.10).

Under three-wave diffraction conditions (Fig. 11---14), the results of computation are very similar to those of the Bragg case. Namely, the intensity of cooperative radiation emitted at large angles lasts much longer than the particle flight time through the crystal, though that is much lower than the radiation intensity emitted in forward direction.

  \begin{figure}[ht]
    \begin{center}
       \resizebox{65mm}{!}{\includegraphics{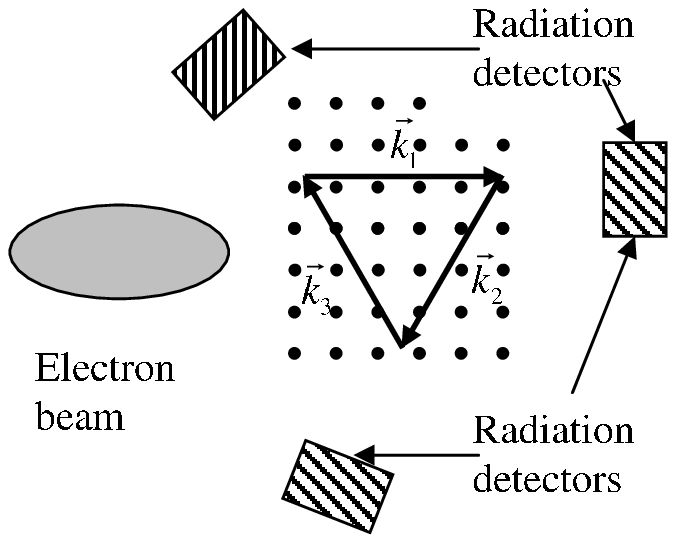}}
    \caption{The three-wave diffraction geometry.}
\label{Fig.11}
    \end{center}
  \end{figure}
  \begin{figure}[ht]
  \begin{center}
       \resizebox{65mm}{!}{\includegraphics{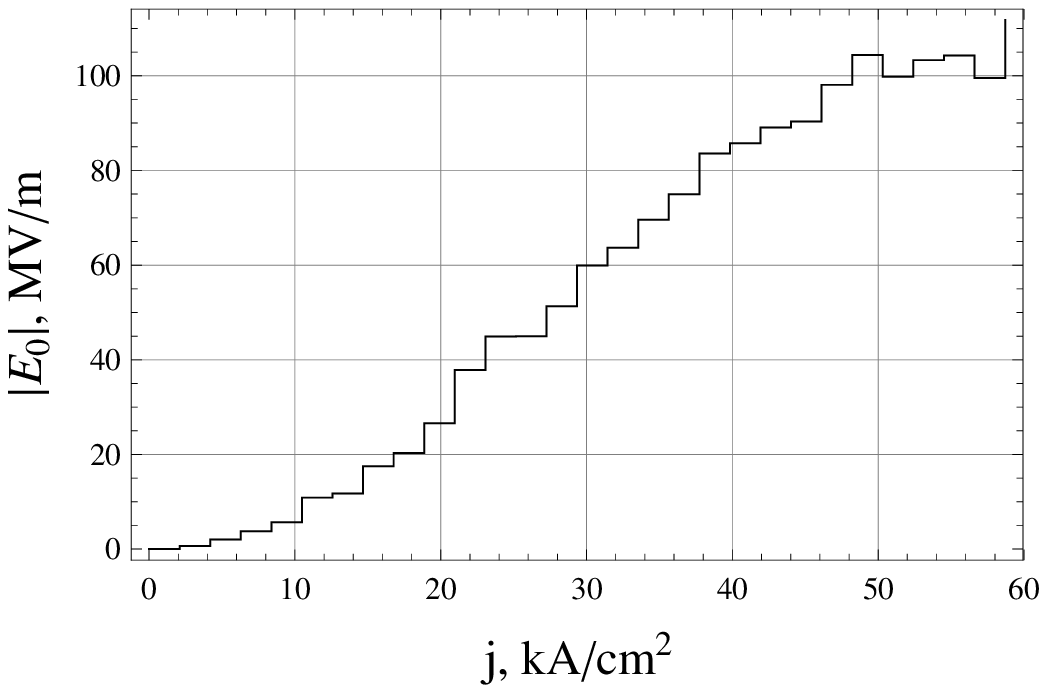}} \resizebox{65mm}{!}{\includegraphics{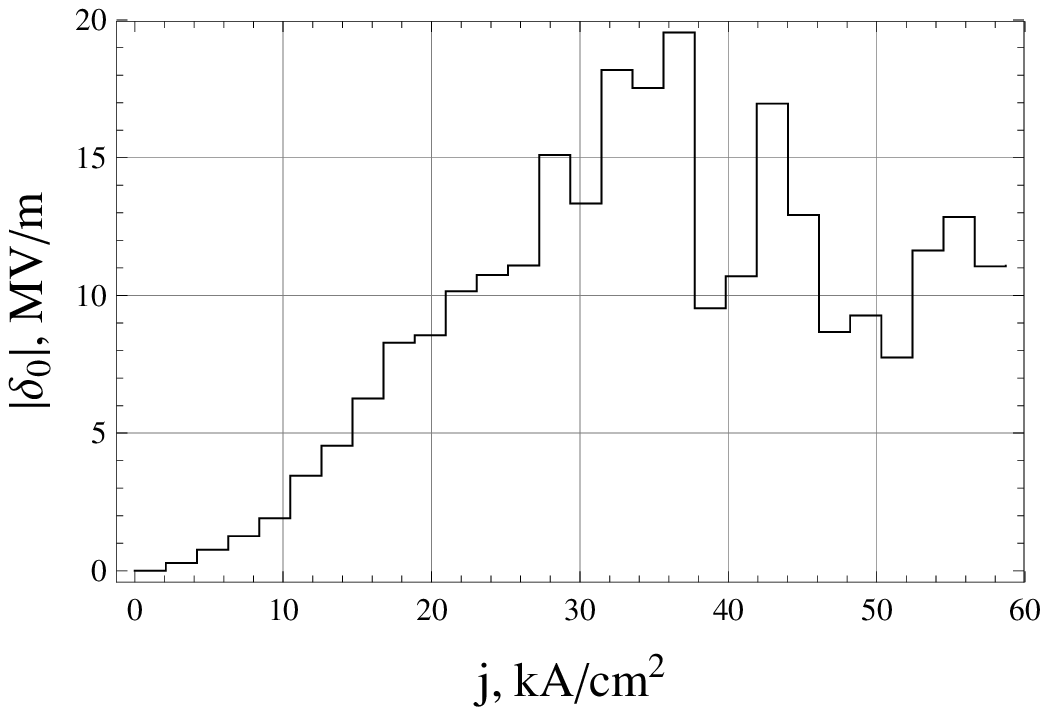}}\\
  \end{center}
    \caption{The three-wave diffraction case. Quasi-Cherenkov radiation at small angles to particles' velocities: radiation amplitude (left), amplitude dispersion (right) [$\nu=0.1$~THz]}
\label{Fig.12}
  \end{figure}
  \begin{figure}[ht]
  \begin{center}
       \resizebox{65mm}{!}{\includegraphics{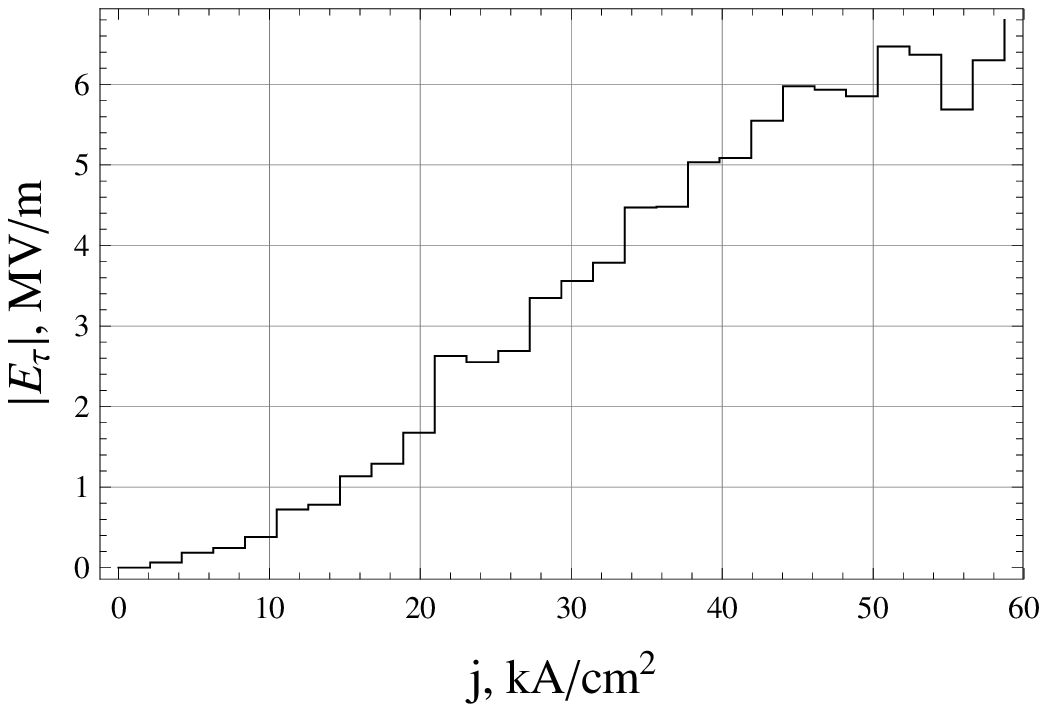}} \resizebox{65mm}{!}{\includegraphics{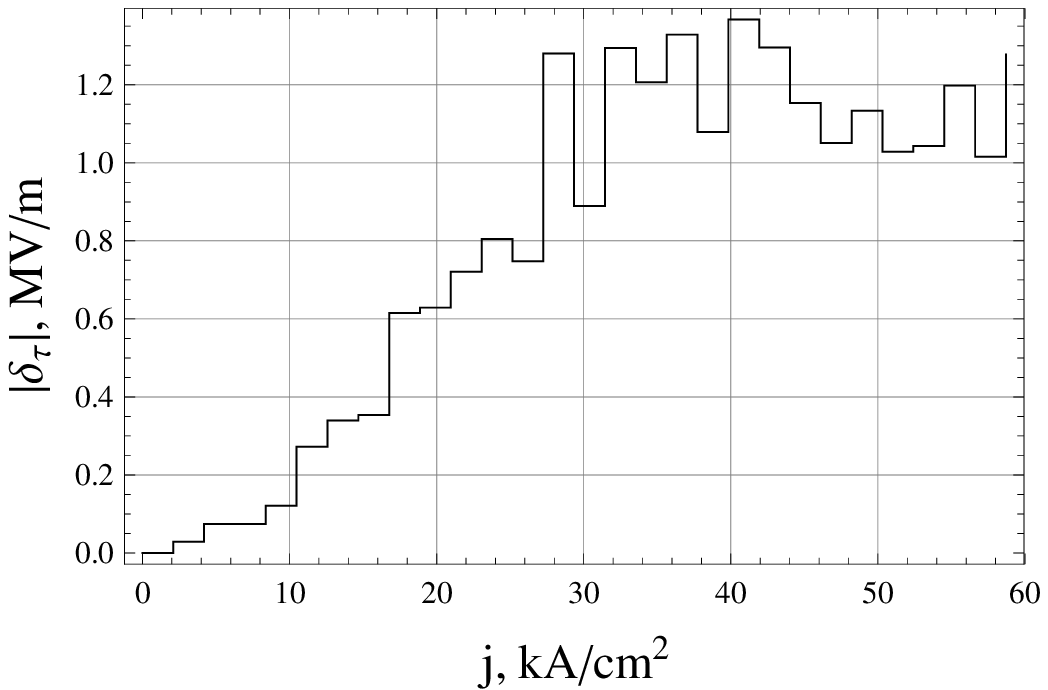}}\\
  \end{center}
    \caption{The three-wave diffraction case. Quasi-Cherenkov radiation at large angles to particles' velocities: radiation amplitude (left), amplitude dispersion (right) [$\nu=0.1$~THz].}
\label{Fig.13}
  \end{figure}
\begin{figure}[ht]
  \begin{center}
       \resizebox{65mm}{!}{\includegraphics{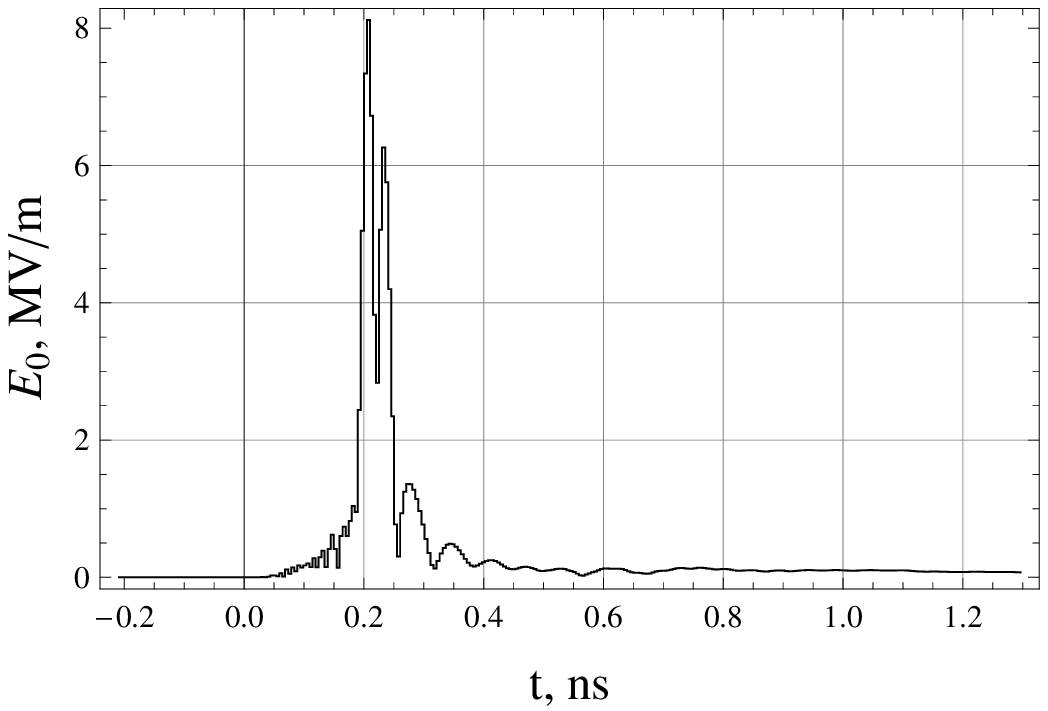}} \resizebox{65mm}{!}{\includegraphics{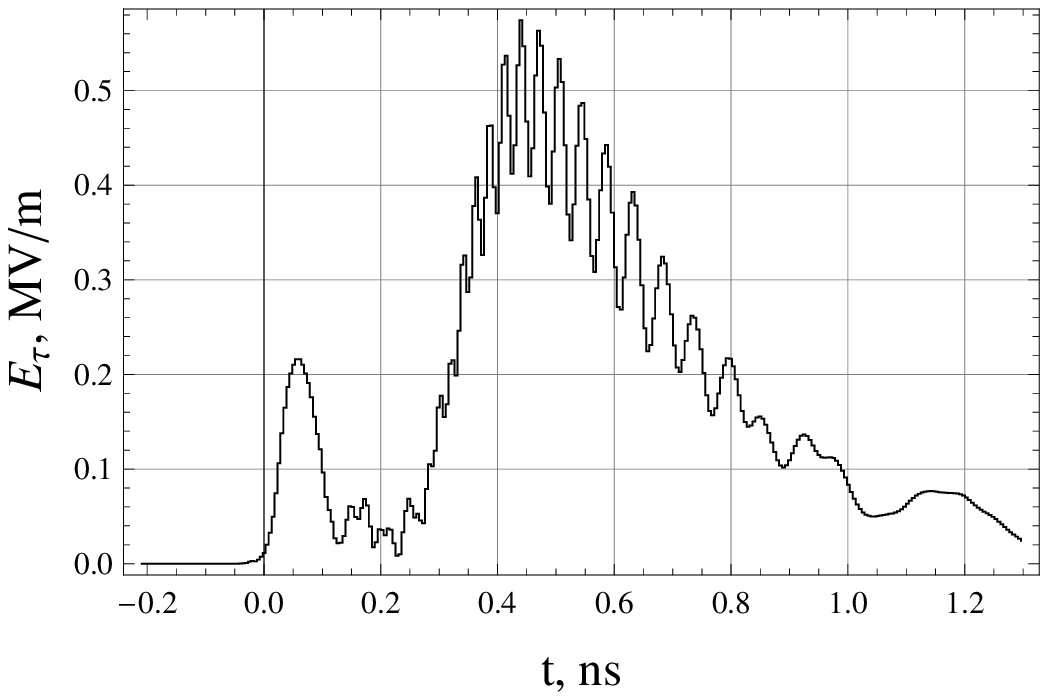}}\\
  \end{center}
    \caption{The three-wave diffraction case. Quasi-Cherenkov radiation at small  (left) and large (right) angles to particles' velocities [$j=10$~kA$/$cm$^2$, $\nu=0.1$~THz]}
\label{Fig.14}
  \end{figure}

\section{Conclusion}
This paper studies the features of parametric (quasi-Cherenkov) cooperative
radiation emitted at both large and small angles to the particle
velocity direction in two- and three-wave diffraction cases. A detailed numerical analysis is given for cooperative THz radiation in artificial crystals. 

The peak intensity of cooperative radiation emitted at small and large angles to particle velocity is investigated as a function of the peak current density. The peak radiation intensity appeared to increase monotonically until saturation is achieved. At saturation, the shot noise causes strong fluctuations in the intensity of cooperative parametric radiation.

It
is shown that, the intensity of cooperative radiation emitted at large angles can last much longer than the particle flight time through the crystal.
At saturation, the shot noise causes strong fluctuations in the intensity of cooperative parametric radiation.
The intensity of THz radiation above 200~MW$/$cm$^2$ is obtained in simulations.

The complicated time structure of cooperative parametric radiation can be observed in artificial (electromagnetic, photonic) crystals in all spectral ranges (X-ray, optical, terahertz, and microwave).

It should be pointed out that thermal fluctuations become
essential if $kT\ge\hbar\omega$ \cite{Anishchenko2014} ($k$ and $\hbar$ is the Bolzman constant and the Plank constant, respectively); namely, when $kT\gg\hbar\omega$,
generation starts as a stimulated emission induced by
thermal quanta rather than as a  spontaneous one. This fact should be taken into account in development of terahertz generators operating at room temperature.

\appendix
\section{Particle-in-cell method}

The set of equations (\ref{Vlasov3}) and (\ref{Maxwell3}) was
solved using the particle-in-cell method, which is widely used in
plasma physics~\cite{Verboncoeur,SveshnikovJakunin1989}. This
method implies that the solution of the kinetic equation  is
modeled using a large number of macroparticles moving along the
characteristics of the kinetic equation. The current and charge
densities are calculated from particle velocities and positions
and are further used for computations of  the electric field on a
space-time mesh. The mesh values of the field are interpolated to
the macroparticle locations; then the forces acting on
macroparticles are calculated. The approach described here is
close to the method described by I.~J.~Morey and C.~K.~Birdsall in
\cite{MoreyBirdsall1989}, which was used for travelling wave tube
modeling.

Let us introduce a spatial  $\omega_z=\{z_n=n\Delta
z,n=0,1,...,n_{max},n_{max}\Delta z=L\}$ and a time
$\omega_t=\{t_s=s\Delta t,s=0,1,...\}$ mesh. Specify an implicit
finite-difference scheme \cite{Kalitkin} of field equations  (\ref{Maxwell3}) with
second order accuracy in time and coordinate (the Bragg case):
\begin{eqnarray}
\label{Maxwell4}
E_{00}^{s+1}=0,E_{\tau nmax}^{s+1}=0,\nonumber\\
0\le n<n_{max}:\nonumber\\
\frac{E_{\tau n+1/2}^{s+1}-E_{\tau n+1/2}^{s}}{c\Delta t}=-\gamma_\tau\frac{E_{\tau n+1}^{s+1/2}-E_{\tau n}^{s}}{\Delta z}-\frac{i\chi_0}{2}E_{\tau n}^{s+1/2}-\frac{i\chi_\tau}{2}E_{0 n+1/2}^{s+1/2},\nonumber\\
0<n\le n_{max}:\nonumber\\
\frac{E_{0n-1/2}^{s+1}-E_{0n-1/2}^{s}}{c\Delta t}=-\gamma_0\frac{E_{0n}^{s+1/2}-E_{0n-1}^{s+1/2}}{\Delta z}-\frac{i\chi_0}{2}E_{0 n-1/2}^{s+1/2}\nonumber\\
-\frac{i\chi_\tau}{2}E_{\tau n-1/2}^{s+1/2}-J_{0n-1/2}^{s+1/2}.
\end{eqnarray}

Let us define the source  $J_{0n}^{s+1/2}$ on the right-hand side
of
 (\ref{Maxwell4}), using the formula
\begin{eqnarray}
\label{Source}
J_{0n}^{s+1/2}=(J_{0n}^{s}+J_{0n}^{s+1})/2,\nonumber\\
J_{0n}^{s+1}=\frac{2\pi\sin\theta e^{i\omega (t+\Delta t)}}{l}\Big(\sum_jQ_j\frac{z_{n+1}-z_j^{s+1}}{\Delta z}e^{-ik_zz_j^{s+1}}\nonumber\\
+\sum_jQ_j\frac{z_j^{s+1}-z_{n-1}}{\Delta z}e^{-ik_zz_j^{s+1}}\Big).
\end{eqnarray}
The contributions to each node come from the particles
concentrated in the domain   $z_{n-1}\le z_j<z_{n+1}$. Summation
in the first and second terms is made over all particles in the
domains $z_n\le z_j^{s+1}<z_{n+1}$ and $z_{n-1}\le
z_j^{s+1}<z_n$, respectively. The weighting factors
$\frac{z_{n+1}-z_j^{s+1}}{\Delta z}$ and
$\frac{z_j^{s+1}-z_{n-1}}{\Delta z}$ are responsible for linear
interpolation of the contributions to the node with number $n$
that come from each particle.

Complete the leap-frog difference scheme (\ref{Maxwell4}) with the
discrete analogues of the equations of motion of macroparticles:
\begin{eqnarray}
\label{EquationOfMotion}
\frac{p_{zj}^{s+1/2}-p_{zj}^{s-1/2}}{\Delta t}=2Q_j\theta Re\big(E_{0j}^{s}e^{ik_zz_j^{s+1/2}-i\omega(t+dt/2)}\big),\nonumber\\
\frac{z_{j}^{s+1}-z_{j}^{s}}{\Delta t}=v_{zj}^{s+1/2},v_{zj}^{s+1/2}=\frac{p_{zj}^{s+1/2}/M_j}{\sqrt{1+(p_{zj}^{s+1/2}/M_jc)^2}}.
\end{eqnarray}
The field $E_{j}^{s}$ at  particles' locations can be
found by means of linear interpolation from surrounding nodes:
\begin{eqnarray}
 \label{Field}
E_{0j}^{s}=\frac{z_{n+1}-z_j^{s}}{\Delta z}E_{0n}^{s}+\frac{z_j^{s}-z_n}{\Delta z}E_{0n+1}^{s},\nonumber\\
z_n\le z_j^{s}<z_{n+1}.
\end{eqnarray}

Injection and extraction of particles are performed as follows:
during every time step, we inject  $\Delta N$ number of particles,
whose initial phases are uniformly distributed on the interval $[0,\pi)$ \cite{Tran}. It should be noted that the quantity $\Delta N$ obeys the Poisson statistics \cite{Neil}
\begin{equation}
 P(\Delta N)=\frac{\Delta N_{av}^{\Delta N}\exp(-\Delta N_{av})}{\Delta N!},
\end{equation}
where $N_{av}$ --- is an average number of particles injected during the time interval $\Delta t$.

\end{document}